\documentclass[11pt]{article} 
\usepackage{amsfonts,amssymb,slashed,makeidx,latexsym,setspace}
\usepackage{graphicx,graphics,amssymb,epsf,rotate,subfigure} 
\begin{document}
\begin{flushright}

\end{flushright}

\vskip 30pt

\begin{center}
{\Large \bf On the possibility of generating leading order gaugino
  masses in direct gauge mediation scenario} \\
\vspace*{1cm} \renewcommand{\thefootnote}{\fnsymbol{footnote}} {{\sf
    Tirtha Sankar Ray $ {}$} } \\
\vspace{10pt} {\small ${}$ {\em Institut de Physique Th\'eorique, CEA-Saclay,\\
F-91191 Gif-sur-Yvette Cedex, France. }}
\normalsize
\end{center}

\begin{abstract} 
Generating gaugino masses at the leading order has typically been
difficult in direct/semi-direct gauge mediated supersymmetry breaking
models. The Komargodski-Shih theorem has established that local
stability of the supersymmetry breaking vacuum implies a vanishing
leading order gaugino mass in generic renormalizable O'Raifeartaigh
models. We relax the condition of renormalizability and investigate
the possibility to evade the KS no-go theorem using higher dimensional
operators in the K\"ahler potential and the superpotential.  We
demonstrate that higher dimensional terms which are polynomial in
superfields are not adequate to evade the KS theorem. We narrow down
on the possible class of non-polynomial corrections that can induce
unsuppressed gaugino mass in a global supersymmetry breaking
vacuum. We find that these models are tantalizingly close to the
theories obtained from strongly coupled supersymmetry breaking
schemes.

\vskip 5pt \noindent
\texttt{PACS Nos: 11.30.Pb 12.60.Jv} \\
\end{abstract}

\setcounter{footnote}{0}
\renewcommand{\thefootnote}{\arabic{footnote}}

\section{Introduction} 
The realization that generalized O'Raifeartaigh (O'R) models of direct
gauge mediated supersymmetry (SUSY) breaking \cite{Giudice:1998bp} are
low energy description of dynamical supersymmetry breaking scenarios
from a strongly coupled sector, has been known for some time now.
Better understanding of this phenomenon was achieved in
\cite{Intriligator:2006dd}, which has kindled renewed interest in
these models. A typically stubborn problem of these scenarios is the
generation of gaugino masses at the leading order, even with explicit
tree level R-symmetry breaking, see \cite{Kitano:2010fa} for a recent
review of direct and semi-direct gauge mediation models. First pointed
out in \cite{Polchinski:1982an}, explicit calculations with all known
renormalizable models of direct gauge mediation have shown that
cancellations lead to zero gaugino masses at the leading order whereas
scalar masses are generally generated at two loop level. Further, the
phenomenon of gaugino mass screening \cite{ArkaniHamed:1998kj}
prevents gaugino masses being generated at the next order in the
messenger loop\footnote{However see \cite{Argurio:2010fn} for ways to
  address this problem by using chiral messengers.}. This further
complicates the possibility to generate sizable gaugino masses in
direct gauge mediation models.  It was finally realized in
\cite{Komargodski:2009jf} that the condition for local stability of
the supersymmetry breaking pseudomoduli direction would prevent
gaugino masses from being generated at the leading order for general
renormalizable models of direct gauge mediation. It was demonstrated
that for stable supersymmetry breaking pseudomoduli direction, the
determinant of the fermionic mass matrix for the messengers is
independent of the pseudomoduli field dependence. This leads to a
vanishing gaugino mass in the leading order which is proportional to,
$ M_g^a \propto \partial \log det(M_f)/\partial X$ where X is the
pseudomoduli field and $M_f$ is the fermionic mass matrix for the
messenger fields.  The vanishing gaugino masses in direct gauge
mediation models is now understood in terms of this Komargodski-Shih
(KS) no-go theorem.

With the early data from the LHC \cite{Chatrchyan:2011zy} constraining
the SUSY spectra in general and the gluino in particular to be
relatively heavy, it has become evermore important to investigate
avenues to generate unsuppressed gaugino masses in direct gauge
mediation models of supersymmetry breaking.  Recently, ways to
ameliorate this problem have been suggested in the literature
\cite{Nakai:2010th} \cite{Maru:2010yx}.  In \cite{Nakai:2010th} the
discussion is based on the fact that the form of the fermionic mass
matrix for the messenger fields is not constrained by the KS theorem
for models with tachyonic directions in the scalar potential. One
would expect leading order gaugino masses to be generated is these
models.  Non-canonical K\"ahler corrections can be used in these
models to lift the tachyonic directions. It has been argued that with
the non-canonical K\"ahler corrections, the effective scalar potential
of these models will not have any tachyonic direction but leading
order gaugino masses will be generated.  In the present paper we make
a complementary investigation. We study the possibility to evade the
KS theorem by introducing non-renormalizable terms to models with
stable supersymmetry breaking vacuum. We consider the possibility that
these contributions introduce a holomorphic pseudomoduli dependence in
the determinant of the fermionic mass matrix for the messengers
generating leading order gaugino masses, without disturbing the vacuum
configuration.
  
We investigate the possibility to generate leading order gaugino
masses by introducing non-renormalizable operators in both the
superpotential and the K\"ahler potential. The most general form of
the non-canonical K\"ahler terms that can contribute to the reduced
fermionic mass matrix of the messenger fields are identified.  We note
that all possible non-renormalizable superpotential terms can be
considered to be a subset of the non-canonical K\"ahler terms as far
as their contribution to the messenger mass matrices in the desired
vacuum is considered. We systematically study the viability of
generating unsuppressed gaugino masses using higher dimensional terms
that are polynomial in the fields. Though we do not specify the UV
completion of these models, they can in principle be considered to
have originated from some perturbative dynamics at higher
energy. However, we find that this class of models are unable to
generate unconstrained gaugino masses which are in general suppressed
by the high cutoff scale $(\Lambda)$ of the effective
non-renormalizable theory.  The lowest order K\"ahler term which
induces nontrivial corrections to the fermionic mass matrix of the
messenger fields has a mass dimension of four.  Qualitatively, we
observe that beyond this order, gaugino masses are suppressed by
$(\langle X \rangle / \Lambda)^{\delta-4}$ where $\langle X \rangle $
is the vev of the pseudomoduli field and $\delta$ is the dimension of
the operator in the K\"ahler potential.  In general one expects
$\langle X \rangle \ll \Lambda,$ hence a large suppression.

Next we relax the condition of perturbative UV completion and consider
more general functions of the fields motivated by strongly coupled
supersymmetry breaking scenarios. We demonstrate that with this
generalization the condition for local stability can be explicitly
solved in the simplest cases. We obtain surprisingly simple solutions
for models of supersymmetry breaking that evade the KS theorem. This
class of models break supersymmetry at the global minimum but
generates unconstrained gaugino masses. However the condition of local
stability of the pseudomoduli direction puts severe constraints on the
functional form of the effective Goldstino-messenger terms in the
superpotential. The general class of interactions that are allowed are
very close to the UV complete theories studied in the literature.

The rest of the paper is organized as follows: In
Section~\ref{basics}, we briefly review the KS theorem within the
renormalizable setup and then lay down the framework to generalize to
non-renormalizable scenarios.  In Section~\ref{rss}, we consider the
possibility to evade the KS theorem using higher dimensional operators
that are polynomial in fields.  In Section~\ref{npc}, we consider the
non-polynomial generalization.  Finally in section \ref{conclusion},
we conclude with some general observations.

\section{Generalization of the KS theorem}
 \label{basics}
\subsection{A review of the KS theorem in the renormalizable scenario}
\label{rkst} 
Consider a general O'R theory with canonical K\"ahler potential and a
renormalizable superpotential.  Let the gauge singlet $X$ and
$\{\phi_a\}$ be a set of chiral superfields which constitutes the
sector that will break SUSY spontaneously.  In order that the
$\{\phi_a\}$ should also act as messengers, they should be charged
under the SM gauge group. The superfield $X$ is an SM singlet and can
get a vev in the vacuum configuration to break SUSY
spontaneously. Typically $X$ represents a flat direction in the scalar
potential.  With this field content, the most general renormalizable
superpotential can be written as, \begin{equation} W= f
  X+{1\over2}(\lambda_{ab}X+m_{ab})\phi_a\phi_b +
  {1\over6}g_{abc}\phi_a\phi_b\phi_c~.  \label{rsp} \end{equation}
Here the fermionic mass matrix for the messenger fields $\{\phi_a\}$
is ${\cal{M}_F}= W_{ab}=\lambda_{ab} X + m_{ab}$, where $W_a \equiv
\partial W/\partial \phi_a .$ In general the determinant of this
matrix may be written as, \begin{equation} det({\cal{M}_F}) =
  det(\lambda_{ab} X + m_{ab}) = \Sigma
  C_n(\lambda,m)X^n.  \label{detfermas} \end{equation} Let the roots
of the polynomial on the RHS of the above equation be defined by
$\Sigma C_n(\lambda,m)X^n|_{X\rightarrow X_0^l}=0$.  At $X=X_0^l$ the
determinant of the fermionic mass matrix vanishes and a Goldstino
direction $(v)$ is defined for every root of the polynomial as
follows, \begin{equation} (\lambda_{ab} X_0^l + m_{ab})v=0.
\label{fermionmass} \end{equation}

The bosonic mass matrix for the messenger fields is given by,
\begin{equation} {\cal{M}}_B^2 = \left( \begin{array}{cc}
\cal{M}_F^*\cal{M}_F & \cal{F}^* \\ \cal{F} & \cal{M}_F\cal{M}_F^*
\end{array} \right), \end{equation} where
${\cal{F}}_{ab}=W_c^*W_{abc}.$ If the pseudomoduli direction is
locally stable everywhere then the scalar mass matrix has to be
positive semidefinite.  However note that if $v$ is the Goldstino
direction defined by ${\cal{M}_F} v =0$ then it is easy to show that,
\begin{eqnarray} \left( \begin{array}{c} v \\ v^* \end{array}
\right)^{\dagger}\left( \begin{array}{cc} \cal{M}_F^*\cal{M}_F &
\cal{F}^* \\ \cal{F} & \cal{M}_F\cal{M}_F^* \end{array} \right) \left(
\begin{array}{c} v \\ v^* \end{array} \right) = v^T {\cal{F}} v + cc.
\end{eqnarray} The RHS of this equation must vanish identically if the
bosonic mass matrix is required to be positive semidefinite, otherwise
one can make the expression negative by rotating the complex phase of
$v$. We conclude that the condition of local stability of the desired
vacuum implies that for a massless Goldstino $(v)$ in the fermionic
sector there exists a flat direction in the scalar potential given by
the vector $(v~~v^*)$.  An important corollary of this is,
\begin{equation} {\cal{F}}_{ab} v=f \lambda_{ab} v = 0.
\label{bosonicmass} \end{equation} Using Eq.~\ref{bosonicmass} in
Eq.~\ref{fermionmass} we find $v$ has to be a simultaneous null
eigenvector of the matrices $\lambda_{ab}$ and $m_{ab}$.  This implies
that $v$ is a null eigenvector of any matrix of the form $\alpha
\lambda_{ab} + \beta m_{ab}.$ It follows that $ det({\cal{M}_F})=0$,
contradicting our original assumption that the determinant is not
identically zero. Thus we find that the assumption taken in
Eq.~\ref{detfermas} is inconsistent and we conclude that,
\begin{equation} det({\cal{M}_F}) = det(\lambda_{ab} X + m_{ab}) =
Const.  \end{equation} It follows that the leading order gaugino
masses given by,
\begin{equation}
M_g^a \sim \frac{\alpha^a}{4\pi}\bar{W}_{\bar{X}}
\frac{\partial}{\partial X} \log det(M_f), \label{gauginomass}
\end{equation}
vanish.  In conclusion 
the KS theorem demonstrates that in renormalizable models of direct gauge
mediation with a locally stable pseudomoduli direction, gaugino masses
are not generated at the leading order.

\subsection{Non-renormalizable generalization} \label{nrg} 
To study this scenario in non-renormalizable set up we first define
the desired vacuum configuration of a theory with the field content of
Section \ref{rkst}. In order to preserve the SM gauge group we should
have $\langle \phi_a \rangle = 0 ~ \forall a$.  The only field that
can take a vev to spontaneously break SUSY is X. Hence we are looking
at a vacuum of the form, \begin{equation} \langle X \rangle
  \rightarrow~ \mbox{undetermined}, ~~~~~~~\{\langle \phi_a \rangle =
  0\} ~~~ \forall a.  \label{vac} \end{equation}

We start with the general renormalizable superpotential given in
Eq.~\ref{rsp}. The superpotential is linear in X representing a flat
pseudomoduli direction in the scalar potential. We find that the two
equations $W_X =W_{\phi_i}=0$ cannot be simultaneously satisfied.  At
the desired vacuum we have $\langle W_{\phi_i}\rangle = 0,\langle
W_X\rangle = f$ and SUSY is broken spontaneously. Considering that the
flat direction is locally stable everywhere the determinant of the
reduced fermionic mass matrix for the messenger fields remains
independent of the pseudomoduli field by the KS theorem implying a
zero gaugino mass at the leading order.  Our objective is to introduce
an $X$ dependence into the determinant of reduced fermionic mass
matrix for the messenger fields by adding non-renormalizable terms to
a theory like this without disturbing the local stability of the SUSY
breaking vacuum.  We will consider non-renormalizable terms both in
the superpotential and in the K\"ahler potential that can generate
such corrections to the mass matrices at the vacuum configuration.

We first consider non-canonical K\"ahler terms.  Following the
notations of \cite{Aldrovandi:2008sc}, the messenger mass matrices for
generic non-canonical K\"ahler potential can be written as,
\begin{equation} {\cal{M}_F}^{NC} = {\cal{M}_F}^C - \Gamma^d_{ab} W_d,
\label{ncfm} \end{equation} where $\Gamma^d_{ab} W_d =(K^{d\bar{e}}
\partial_a K_{b\bar{e}}) W_d$. The bosonic mass matrix also receives
further corrections due to the non-canonical Kahler terms and can be
given as, \begin{equation} ({\cal{M}}^{NC}_B)^2 = \left(
\begin{array}{cc} {{\cal{M_F}}}^{NC} {\cal{M_F}^*}^{NC} -
\bar{W}_{\bar{a}}(R_{\bar{b}b})^{a\bar{a}} W_a & {{\cal{F}}}^{*NC} \\
{{\cal{F}}}^{NC} & {{\cal{M_F}}^*}^{NC} {\cal{M_F}}^{NC} -
\bar{W}_{\bar{a}}(R_{\bar{b}b})^{a\bar{a}} W_a \end{array} \right),
\label{bosonmm} \end{equation} where $\bar{W}_{\bar{a}}(R_{\bar{b}b})^{a\bar{a}}
W_a = \bar{W}_{\bar{a}}(K^{\bar{a}c} \partial_{\bar{b}}
\Gamma^a_{bc})W_a$ and ${{\cal{F}}}^{NC} =
\partial_{bc}(W_{a}K ^{\bar{a}a})\bar{W}_{\bar{a}}$ .  Considering
that in the vacuum we can only have $W_X = \bar{W}_{\bar{X}} \neq 0,$
the nonzero components are given by, $(R_{\bar{b}b})^{X\bar{X}} \sim
K^{\bar{X}c} \partial_{\bar{b}} \Gamma^X_{bc}$ and ${{\cal{F}}}^{NC}
\sim \partial_{bc}(W_{a}K ^{\bar{X}a})\bar{W}_{\bar{X}}$ .

By inspecting Eq.~\ref{ncfm}, one can see that the new terms need to
be bilinear and holomorphic in the messenger fields in order to
contribute to the fermionic messenger mass matrices.  Thus the most
general structure of the non-canonical part of the K\"ahler potential
that contributes to the fermionic mass matrices of the messenger
fields may be symbolically represented as, \begin{equation} K \supset
  C_{ab} \phi_a \phi_b f(\frac{X}{\Lambda},\frac{\bar{X}}{\Lambda}) +
  cc, \label{nckt} \end{equation} where $C_{ab} \neq 0 \Leftrightarrow
Q(\phi_a\phi_b) = 0$ and all other terms are zero. $Q(\hat{O})$
represents all the charges of the operator \^{O} under the SM gauge
groups.

With this form of the K\"ahler terms the curvature tensor
$\bar{W}_{\bar{a}} (R_{\bar{b}b})^{a\bar{a}} W_a = 0$.  We note that
the presence of a non zero curvature tensor in the K\"ahler metric
results in new contribution to the gaugino masses. With these new
contributions it is impossible to recast the scalar and fermionic
messenger mass matrices in the form, \begin{eqnarray}
  W_{eff}^{mess}&=&M_{ab}\phi_a\tilde{\phi_b} + \theta^2
  F_{ab}\phi_a\tilde{\phi_b}, \nonumber
  \\ {\cal{L}}_{eff}^{mess}&=&-(M_{ab}\psi_a\bar{\psi_b}+h.c.)
  -(\varphi_a\tilde{\varphi_a}^*) \left( \begin{array}{cc}
    MM^{\dagger} & F^*\\ F & M^{\dagger}M \end{array}
  \right)\left( \begin{array}{c}
    \varphi_b^*\\ \tilde{\varphi_b} \end{array}
  \right).  \end{eqnarray} where $\psi$ and $\varphi$ are the
fermionic and scalar component respectively, of the chiral messenger
superfield $\phi.$ This would potentially cause the generated gaugino
masses to deviate from the expression given in
Eq.~\ref{gauginomass}. This in itself is an interesting avenue to
generate leading order gaugino masses in direct gauge mediation models
and needs to be explored further.  However the arguments of the KS
theorem crucially depend on the expression for the gaugino masses as
given by Eq.~\ref{gauginomass} and are not well understood in
scenarios where this is no longer true.  In this paper we will be
confined to models where the curvature tensor identically vanishes.
With this choice the only new contributions to the mass matrices are
given by, \begin{eqnarray} {\cal{M_F}}^{NC} &=& {\cal{M_F}}^{C}
  -C_{ab} \langle W_X\rangle f_X(\frac{X}{\Lambda},\frac{\bar
    X}{{\Lambda}}), \label{fmm} \\ {\cal{F}}^{NC} &=& {\cal{F}}^C
  -C_{ab} |W_X|^2 f_{X\bar X}(\frac{X}{\Lambda},\frac{\bar
    X}{{\Lambda}}), \label{NCFT}
\end{eqnarray} where $ f_{x} \equiv \partial f / \partial x.$

At this stage we note that the arguments for the KS theorem used in
the canonical case are no longer applicable. We find that if $v$ is
now a simultaneous eigenvector of both ${\cal{F}}^{NC}$ and
${\cal{M_F}}^{NC}$ one cannot argue that the determinant of
${\cal{M_F}}^{NC}$ has to be identically zero everywhere. This is
because the matrix form of ${\cal{F}}^{NC}$ is in general different
from ${\cal{M_F}}^{NC}$. They also have different dependences on
$X~\mbox{and/or}~\bar X$.  Some generic observations are now in order:
\begin{itemize} \item The KS argument is valid only in case of a
locally stable pseudomoduli directions i.e., for scenarios where the
reduced scalar messenger mass matrix is positive semidefinite. The
assertion that new contributions from the non-canonical Kahler
potentials can evade this argument and generate leading order gaugino
masses should be supplemented by an example by example demonstration
that these additional terms should not destabilize the scalar mass
matrix.  \item Corrections to the K\"ahler terms can potentially lead
to wrong sign kinetic terms in certain region of the field space. And
this consideration puts stringent constraints on the possible form of
higher dimensional corrections that are allowed in the K\"ahler
potential. However one can assume that high energy dynamics near the
cutoff scale can fix this malady. We will ignore this consideration
with the understanding that cutoff scale is much larger than the scale
of SUSY breaking.  \end{itemize}

We now turn our attention to possible non-renormalizable
superpotential terms.  The most general superpotential term that
contributes to the fermionic mass matrix for the messenger fields, in
the vacuum configuration defined in Eq.~\ref{vac} is given by,
\begin{equation} \Delta W_{NR} = m \phi_{a} \phi_{b}
g(\frac{X}{\Lambda}).  \label{nrsp} \end{equation} The contribution of
this term to the mass matrices in the desired vacuum configuration is
identical to the K\"ahler potential given in Eq.~\ref{nckt} with the
following identifications, \begin{equation}
f(\frac{X}{\Lambda},\frac{\bar{X}}{{\Lambda}}) =
\frac{\bar{X}}{{\Lambda}} g(\frac{X}{\Lambda}) ~~\mbox{and}~~ m~~ =
\frac{C_{ab}\langle W_X \rangle}{\Lambda}.  \label{wkcor}
\end{equation} Thus we note that the most general non-renormalizable
terms that can be added to the superpotential and can contribute to
the mass matrices are a specific subset of the most general
non-canonical K\"ahler terms as far as their contribution in the
vacuum configuration is considered. It follows that a study of the
effect of non-renormalizable terms in direct gauge mediation models
can be effectively carried out by considering the non-canonical terms
in the K\"ahler potential alone.

Having made this observation it should be noted that there are
definite differences between a higher dimensional superpotential term
and a non-canonical K\"ahler term.  These differences show up in the
global structure of the scalar potential specifically in the field
space regions away from the SUSY breaking vacuum.

\section{Theories with polynomial corrections} \label{rss} If we
consider a perturbative UV completion of the theories, we can expect
these effective terms to be generated by integrating out heavy states
operative at high scale.  This consideration constraints the
functional form of $f$ defined in Eq.~\ref{nckt} and Eq.~\ref{nrsp} to
be a polynomial of the fields.  In this section we will discuss the
possibility of evading the KS theorem to generate unconstrained
gaugino masses using such polynomial correction to the K\"ahler
potential and the superpotential.

\subsubsection*{Non-Canonical K\"ahler potentials:} 
Let us consider that the function $f$ in Eq.~\ref{nckt} is a
polynomial in both $X$ and $\bar X$. Thus generically we may
write, \begin{equation} f(\frac{X}{\Lambda},\frac{\bar{X}}{{\Lambda}})
  = \sum_{n\bar{n}} C^{n\bar{n}} \frac{X^n\bar{X}^{\bar
      n}}{\Lambda^{n+\bar{n}}}.
\label{polyncK} \end{equation} In this case the contributions to the
matrices are of the following form, \begin{eqnarray} {\cal{M_F}}^{NC}
  &=& {\cal{M_F}}^{C} -\sum_{n\bar{n}} C^{n\bar{n}}_{ab} \langle
  W_X\rangle \bar{n} \frac{X^n\bar{X}^{\bar n
      -1}}{\Lambda^{n+\bar{n}}},
\label{fmm1}\\ {\cal{F}}^{NC} &=&{\cal{F}}^C -\sum_{n\bar{n}}
C^{n\bar{n}}_{ab}|W_X|^2 \bar{n}n \frac{X^{n-1}\bar{X}^{\bar n
    -1}}{\Lambda^{n+\bar{n}}}.  \label{NCFT1} \end{eqnarray}

It is clear from Eq.~\ref{fmm1} and Eq.~\ref{NCFT1} that for the new
non-renormalizable terms to contribute we should ensure $\bar{n} \neq
0.$ We will now summarize how the individual terms contribute to the
gaugino mass and the stability condition for various choices of
$n,\bar{n}$.  \begin{itemize} \item The lowest order contribution
  comes from the term $\bar{n}=1,~n=0.$ In this case we find that the
  new contribution is just a redefinition of the matrix
  $m_{ab}\rightarrow m_{ab} -C_{ab} \langle W_X\rangle /\Lambda$. We
  can now trace the arguments given in Section~\ref{rkst}
  identically. This will naturally lead to the conclusion that if the
  vacuum is locally stable, leading order gaugino masses will
  vanish.  \item The next order contribution comes when $n=1,~\bar n
  =1.$ In this case we find that the contribution simply results in a
  redefinition of the matrix $\lambda_{ab} \rightarrow \lambda_{ab} -
  C_{ab} \langle W_X\rangle/\Lambda^2$. This again leads to the same
  conclusion as in the previous case.  \item At this same order we
  have a non-trivial contribution given by $\bar{n}=2,~n=0$. This
  contributes to the fermionic mass matrix but does not contribute to
  $\cal{F}$. This cannot be modeled by redefinition of
  parameters. However we make the observation that this term cannot
  directly introduce a holomorphic dependence on $X$, into the
  fermionic mass matrix.  With the observation that $det(\lambda_{ab}
  X +m_{ab})= Const,$ we expect the $det(\lambda_{ab} X +m_{ab}-
  C_{ab} \langle W_X\rangle \bar{X}/\Lambda^2 ) \sim
  \bar{X}X/\Lambda^2. $ This will lead to gaugino mass terms that are
  suppressed by the factor $\langle \bar{X} \rangle/\Lambda$. In
  general it is well known that in O'R models the one loop correction
  fixes the $X$ vev near zero \cite{Dudas:2010qg}. This will certainly
  be modified due to the presence of the non-canonical K\"ahler
  terms. It is still expected that the vev will be generally at a
  scale where $\langle X \rangle \ll \Lambda$ and thus lead to a
  suppression of the generated gaugino masses.  \item All higher order
  non-canonical K\"ahler terms with $\bar{n}+n>2$ will in general lead
  to further suppression in the gaugino mass terms of the order
  $\left(\frac{\langle X
    \rangle}{\Lambda}\right)^{n-1}\left(\frac{\langle \bar{X}
    \rangle}{\Lambda}\right)^{\bar{n}-1}$.  \end{itemize} In
conclusion we observe the generic non-canonical K\"ahler terms of
perturbative origin when added to O'R models with global SUSY breaking
can only lead to leading order gaugino masses which are suppressed by
the cutoff scale.  This general observation is made without any
reference to the stability condition of the vacuum.  Note that in this
class of models the determinant of the fermionic mass matrix will be a
polynomial in $X$ and therefore will have roots in the finite complex
plane.  The pseudomoduli direction will in general have an instability
at the point where the determinant vanishes.

\subsubsection*{Non-Renormalizable Superpotential terms:} In
continuation of the discussion in the previous section we point out
that the most general non-renormalizable terms in the superpotential
which are polynomial in the superfields are a subset of the K\"ahler
potential defined in Eq.~\ref{polyncK}.  In the phenomenologically
acceptable vacuum, the contribution to the messenger mass matrices
from these non-canonical K\"ahler terms with $\bar{n}=1$ corresponds
to the contribution from the most general non-renormalizable
superpotential term given by, \begin{equation} \Delta W = \sum_n
  m^{(n)}_{ab} \phi_a\phi_b \left(\frac{X}{\Lambda}\right)^{n},
\end{equation} where $m^{(n)}$ can be read off from Eq.~\ref{wkcor}.
The limitations of such terms for $n=0,1,>1$ are similar to the ones
discussed earlier.

We make the general observation that starting with a direct gauge
mediation theory where SUSY is broken globally and the leading order
gaugino masses disappear due to the KS theorem, it is impossible to
generate them by adding non-renormalizable terms that are polynomial
in the fields, either to the superpotential or the K\"ahler potential.

\section{Theories with non-polynomial correction }\label{npc} With the
conclusion of the previous section we abandon the possibility of
circumventing the KS theorem using higher dimensional terms that are
polynomial in the superfields, possibly arising from perturbative
dynamics at high energy scales. Instead we turn our attention to terms
arising from theories with non-perturbative UV completion.  Effective
low energy description of non-perturbative theories of SUSY breaking
can give rise to terms that are non-polynomial in the superfields.
The theories of dynamical SUSY breaking \cite{Witten:1981nf},
\cite{Dine:1981gu}, commonly incorporate terms that are exponential of
the superfields. In theories where gaugino condensates are utilized to
break SUSY, the exponential of the dilaton fields commonly appears
\cite{Dine:2010cv}. In retrofitted O'R models \cite{Dine:2006gm} where
the vev of the pseudomoduli is dynamically generated, we find the
effective superpotential at energies below the dynamical scale
contains terms where the pseudomoduli superfields appear in the
exponential.  Non-polynomial terms arise in the effective
superpotential of SUSY theories with ISS type supersymmetry
breaking. This is essentially generated from the dual of
non-perturbative strongly coupled SQCD like theories
\cite{Seiberg:2008qj} \cite{Intriligator:2007cp}.  In this class of
theories the pseudomoduli field commonly appears with negative powers
in the superpotential and the K\"ahler potential.  In the present
paper our paradigm is to take a bottom up approach to the problem of
generating leading order gaugino masses in the O'R models, thus
evading the KS theorem. We will neither endeavor to construct a UV
complete theory of the hidden sector nor try to demonstrate the
ability to evade the KS theorem with non-polynomial terms in complete
generality. Rather our approach will be to investigate this as a
possibility using examples.

To keep matters simple we will look at the possibility of adding a
non-renormalizable superpotential term to theories that break
supersymmetry globally.  We will consider the simplest supersymmetry
breaking scenario.  Let $X$ be the Standard Model gauge singlet chiral
superfield.  And $(\phi ~\tilde{\phi})$ is a vector-like\footnote{
  These charged messenger superfields can be considered to fill a
  complete representation of a GUT gauge group like the SU(5) required
  to preserve gauge coupling unification.} pair of messenger fields
charged under the Standard Model gauge group.  The simplest SUSY
breaking sector that can be constructed with this field content is
given by the following superpotential, \begin{eqnarray} W &=& -\mu^2 X
  +f(X) \phi \tilde{\phi}.  \label{no1} \end{eqnarray} We will assume
the that the K\"ahler potential is canonical.  The condition that the
theory generates non-zero gaugino mass at leading order means that
$f(X)$ has to be a non-constant function of $X$.  If we further demand
that the theory breaks supersymmetry globally, one needs to impose the
condition $f(X) \neq 0$ everywhere in the finite complex plane.  Note
that this condition is far stronger than the requirement of local
stability which is enough to discuss the KS theorem.

If we insist that the superpotential is holomorphic in the entire
complex plane then $f(X)$ should also be an analytic function of $X.$
This implies that $f(X)$ is an entire function and subject to the
constraints of the Little Picard theorem.  The examples of entire
functions that do not take the value of zero in the entire finite
complex plane are limited.  From a phenomenological perspective a well
motivated choice would be to take $f(X) = m e^{-\frac{X}{\Lambda}}$ in
Eq.~\ref{no1}. This is the simplest entire function that is non-zero
everywhere in the finite complex plane.  Thus we expect SUSY to be
broken globally in this model.  In the desired vacuum the mass
matrices for the messenger fields now take the following form,
\begin{eqnarray} m_f = me^{-\frac{X}{\Lambda}} ~\mbox{and} ~ m_B^2 =
\left( \begin{array}{cc} m^2 e^{-\frac{X + X^*}{\Lambda}} &~ \frac{m
\mu^2}{\Lambda} e^{-\frac{X^*}{\Lambda}}\\ \frac{m \mu^2}{\Lambda}
e^{-\frac{X}{\Lambda}} &~ m^2 e^{-\frac{X + X^*}{\Lambda}} \end{array}
\right).  \end{eqnarray} The condition for local stability of the
pseudomoduli direction now reduces to, \begin{equation} |m^2
e^{-\frac{X + X^*}{\Lambda}}| < |\frac{m \mu^2}{\Lambda}
e^{-\frac{X}{\Lambda}}|.  \end{equation} As is evident, this condition
is easily violated at finite values of $X$, rendering the vacuum
unstable at that point.  Typically, these instabilities leads to a
vacuum with anomalous breaking of the Standard Model gauge group.  It
should be noted that this conclusion is not an artifact of the simple
form of the superpotential considered and it cannot be resolved by a
simple enlargement of the messenger sector.

\subsection{Generic solution to the local stability condition}
Finally, we abandon the constraint that $f(X)$ is analytic
everywhere. Rather we directly try to solve for condition of local
stability. Using Eq.~\ref{no1}, the scalar mass matrix for the
messenger fields is given by, \begin{equation} m_B^2 = \left(
\begin{array}{cc} |f(X)|^2 &~ -\left(\mu^2 \frac{\partial
f(X)}{\partial X} \right)^*\\ -\mu^2 \frac{\partial f(X)}{\partial X}
  &|f(X)|^2 \end{array} \right).  \end{equation} To establish that a
$2 \times 2$ matrix is positive definite it is enough to show that the
trace and the determinant are positive. The condition on the trace is
trivially satisfied by the above matrix. We turn our attention to the
determinant. The condition that the determinant has to be positive
implies, \begin{equation} |f(X)|^4 \ge \left|\mu^2 \frac{\partial
    f(X)}{\partial X}\right|^2.  \label{bounde2} \end{equation} We
consider the scenario that saturates this bound. To solve the
resulting equation we separate the real and the complex parts, giving
the relation, \begin{equation} \frac{f(X)^2}{\mu^2 ~\partial
    f(X)/\partial X} = \left( \frac{f(X)^2}{\mu^2 ~\partial
    f(X)/\partial X}\right)^* = e^{i\theta}.  \end{equation} This
simplifies to the following differential equation, \begin{equation}
  f(X)^2 = e^{i\theta} \mu^2 \frac{\partial f(X)}{\partial
    X}.  \end{equation} The functional form of $f(X)$ can be easily
obtained by solving the differential equation which gives
us, \begin{equation} f(X) = \frac{\mu^2 e^{i\theta}}{X +
    b}.  \end{equation} Note that this solution saturates the bound
given in Eq.~\ref{bounde2}.  Without any loss of generality we can
choose the function to be $f(X) = m^2/X$, where $m$ is a real
constant.  We observe $f(X)$ though not defined at $X=0$, is analytic
everywhere else.  As long as $ \langle X \rangle \neq 0$, the theory
defined by the superpotential given in Eq.~\ref{no1} is well behaved.
To demonstrate the local stability of this theory we consider the
scalar mass matrix which now takes the following
form, \begin{equation} m_B^2 = \left( \begin{array}{cc}
    \frac{m^4}{|X|^2} &~ \frac{m^2\mu^2}{(X^*)^2}
    \\ \frac{m^2\mu^2}{X^2} &~ \frac{m^4}{|X|^2} \end{array} \right).
\end{equation} We note that the eigenvalues of this matrix are given
by $(m^2 - \mu^2)m^2/|X|^2$ and $(m^2 + \mu^2)m^2/|X|^2.$ Thus, for
$m^2 > \mu^2$, the eigenvalues are positive for any value of $\langle
X \rangle$ and matrix is positive definite. Therefore with this
constraint on the parameters the pseudomoduli direction is locally
stable everywhere.  Importantly, we also note that $f(X)$ does not
take the value zero in the finite complex plane.  This means that not
only the pseudomoduli direction is locally stable everywhere,
supersymmetry is also broken globally. It naturally satisfies all the
conditions we laid down on $f(X)$ at the beginning of this section.
Let us now investigate the global structure of the scalar potential.
The potential $V=\sum_a W_a$ where,
 \begin{eqnarray} W_X &=&
  -\mu^2 -m^2\phi\tilde{\phi}/X^2, \nonumber \\ W_{\phi} &=&
  m^2\tilde{\phi}/X,\\ W_{\tilde{\phi}} &=& m^2\phi/X. \nonumber
\end{eqnarray} Clearly these three equations cannot be simultaneously 
put to zero and supersymmetry is broken globally.  Curiously the
condition $m^2 > \mu^2$ implies that there is only one global 
minimum\footnote{A lower lying minimum only appears when $m^2 <
\mu^2$, in this case the minimum is at $V=(\mu^2 - m^2)m^2$.}
of the potential given by $\langle X \rangle\rightarrow
~\mbox{undetermined and }~ \langle \phi \rangle = \langle \tilde{\phi}
\rangle =0$ and $V=\mu^4$.  With the single
constraint on the superpotential parameters, we not only ensure that
the desired vacuum is locally stable but also enforce it to be the
global minimum of the scalar potential.

The fermionic mass matrix for the messenger is simply given by,
\begin{equation} det(m_f) = m^2/X.  \label{detme2} \end{equation}
Gaugino masses are generated at the leading order. Using
Eq.~\ref{gauginomass} and Eq.~\ref{detme2} we find that,
\begin{equation} M_a \sim \frac{\alpha_a}{4 \pi} \mu^2
\frac{1}{ \langle X \rangle}, \end{equation} which is unsuppressed by
any high scale.  And unlike the minimal gauge mediation models, within
this framework the messenger masses may be in the TeV scale and
observable at the present collider experiments.  This will potentially
lead to interesting phenomenological scenarios at collider
experiments.

In conclusion we note that the possibility to generate gaugino masses
at leading order through direct gauge mediation with locally stable
SUSY breaking vacuum is restricted to very specific class of models
even in its non-perturbative generalization.  Crucially the
interactions of the pseudomoduli field with the messengers are
restricted to have very specific functional forms.  This brings us to
the possible origin of this class of superpotentials.  It is well
known that models of supersymmetry breaking with an SQCD sector
generate effective superpotentials at low energies which have the
pseudomoduli fields appearing in the denominator
\cite{Intriligator:2007cp}. However, we could not find an instance in
the literature where the effective term discussed here appears in its
exact form.  To the best of our knowledge, such terms can not be
generated within the framework of the simplest non-perturbative
scenarios like the ISS.

\section{Conclusion} \label{conclusion} In this paper we have studied
the possibility of adding simple non-renormalizable terms to globally
stable SUSY breaking O'R models to evade the KS no-go theorem. This is
complementary to the study carried out in \cite{Nakai:2010th} where
unstable renormalizable theories were considered and non-canonical
K\"ahler terms were used to lift these instabilities.

Within this framework we have demonstrated that the simple higher
dimensional terms which are polynomial in the fields, and thus can
potentially be generated through perturbative dynamics at higher
scales, are not adequate to alleviate the problem of generating large
unconstrained gaugino masses.  Typically we find in these models the
gaugino masses are suppressed by the high cutoff scale of the
effective theory. Further they exhibit tachyonic directions along the
pseudomoduli direction at points where the determinant of the
fermionic mass matrix vanishes.

Next we have considered non-polynomial terms that can generate
unconstrained gaugino masses without disturbing the stability of the
vacuum.  In this context we have imposed a stronger constraint on the
theory, demanding that the desired SUSY breaking vacuum is the global
minimum of the scalar potential. With these restrictive constraints we
solved for the condition of local stability of the potential.  We
obtain a surprisingly simple solution that satisfies all the
conditions of local and global stability and generates unsuppressed
gaugino masses at the leading order.  We observe that supersymmetry
breaking models having these virtues will have a very specific form of
superpotential where the pseudomoduli field couples to messenger field
with inverse one power.  This might have consequences for Goldstino
couplings and can have major cosmological impact. A systematic
discussion of these issues is beyond the mandate of this paper.  The
form of the non-polynomial terms required for this is also
tantalizingly close to the ones that originate from generic
non-perturbative schemes of SUSY breaking discussed in the literature.

A more thorough study of possible non-polynomial terms described in
the literature should be carried out in the context of direct gauge
mediation models. The possibility of using them to evade the KS
theorem and generate phenomenologically viable soft SUSY breaking
spectrum needs to be carried out. In this context we also note that
the entire discussion in this paper is carried out within a framework
where the K\"ahler metric is flat i.e., the curvature tensor is
considered to be zero everywhere. Relaxation of this constraint may
lead to more phenomenologically acceptable avenues to evade the KS
theorem.

\vskip 5pt

\noindent {\bf{Acknowledgments:}}~ This work would not have been
possible without the support and guidance of St\'{e}phane Lavignac.
The work of TSR is supported by EU ITN , contract "UNILHC"
PITN-GA-2009-237920, the CEA-Eurotalent program and the Agence
Nationale de la Recherche under contract ANR 2010 BLANC 0413 01.

\end{document}